# A Hierarchical Dirichlet Process Model with Multiple Levels of Clustering for Human EEG Seizure Modeling

**Drausin Wulsin**                                                                                          WULSIN@SEAS.UPENN.EDU
Dept. of Bioengineering, University of Pennsylvania, Philadelphia, PA USA

**Shane Jensen**                                                                                           STJENSEN@WHARTON.UPENN.EDU
Dept. of Statistics, University of Pennsylvania, Phildelphia, PA USA

**Brian Litt**                                                                                             LITTB@MAIL.MED.UPENN.EDU
Depts. of Neurology and Bioengineering, University of Pennsylvania, Philadelphia, PA USA

## Abstract

Driven by the multi-level structure of human intracranial electroencephalogram (iEEG) recordings of epileptic seizures, we introduce a new variant of a hierarchical Dirichlet Process—the multi-level clustering hierarchical Dirichlet Process (MLC-HDP)—that simultaneously clusters datasets on multiple levels. Our seizure dataset contains brain activity recorded in typically more than a hundred individual channels for each seizure of each patient. The MLC-HDP model clusters over channels-types, seizure-types, and patient-types simultaneously. We describe this model and its implementation in detail. We also present the results of a simulation study comparing the MLC-HDP to a similar model, the Nested Dirichlet Process and finally demonstrate the MLC-HDP's use in modeling seizures across multiple patients. We find the MLC-HDP's clustering to be comparable to independent human physician clusterings. To our knowledge, the MLC-HDP model is the first in the epilepsy literature capable of clustering seizures within and between patients.

## 1. Introduction

Our work is motivated by the structure of clinical intracranial electroencephalogram (iEEG) in the context of evaluating epilepsy patients for resective brain surgery. Consider a set of patients, each of whom has a number of seizures while being recorded in a hospital's epilepsy monitoring unit. The number of recorded seizures can range widely from only one to over fifty for a given patient, and the iEEG of each is defined by the activity of each recorded electrode channel. The number of channels and their placement can range widely from patient to patient, often with 100-200 individual channels per patient.

Current clinical practice involves clinicians examining the dynamics of seizures to ascertain important clinical factors like how similar (or dissimilar) an individual patient's seizures are to each other. Such information helps the physicians identify which areas of the brain to remove. Almost all steps of this decision process are currently manual, performed by neurophysiology physicians, whose training and decision processes can vary greatly. One might say that this work up to epilepsy surgery is still quite "messy," which may help explain the mediocre outcomes of this surgery for extra-temporal lobe surgeries (de Tisi 2011).

Statistical models can offer decision support for clinical questions like "what are the types of seizures that a patient has" and "which other patients is this patient similar to." A challenge of the data is that every seizure of every patient is unique, though there are similarities between seizures and patients. Currently, most approaches create models in space or time of a single seizure.

This approach is not at all similar to a physician's when analyzing a seizure. A physician appreciates that the dynamics of each seizure are unique but does not forget about the other seizures of that patient or even the other seizures of other patients. The other seizures inform the physician's understanding of the current seizure in some way or another. We believe that hierarchical Bayesian models are a good class of models for this problem since they allow for individual models over local data while still allowing global data to have some influence in the model. "Nonparametric" Bayesian approaches are also attractive because they reduce the amount of necessary model selection.

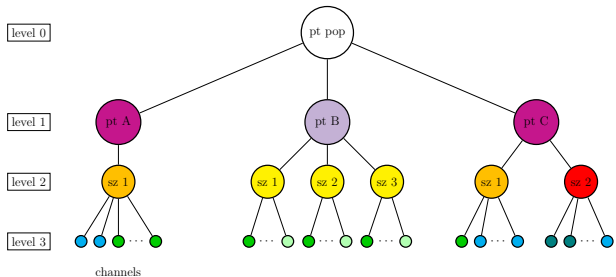

Figure 1. A schematic of the epilepsy data over a population of patients (level 0), each of whom (level 1) has a variable number of seizures (level 2) and a variable number and unique placement of electrode channels (level 3). We assume that the data cluster at each level, as denoted by the colors.

Comparing seizures between patients is challenging because the number and placement of the iEEG channels is unique for every patient[1]. One approach to this problem is to use features of the data that generalize across an arbitrary number of channels. For example, (Schiff et al. 2005) present six intuitive features that seem to capture global properties of a seizure. While this approach is attractive because it is so straightforward, we believe that it is likely to miss the important dynamics that can occur in just a few channels out of a hundred. In many cases, it is just a few channels that are of most clinical interest to physicians.

An alternative approach to this problem is to treat each channel as an i.i.d. sample from an underlying distribution over the space of channel dynamics. In this setting, the number of channels actually present becomes less important, as they can just be thought of as observations from a channel distribution. Due to the complexity of neurophysiologic activity during a seizures, the distribution of channel behaviors will almost certainly be multi-modal. Mixture models are a straightforward approach to such density estimation, and in the case where our channel-features are continuous, we may use the familiar Gaussian Mixture Model (GMM).

We thus are interested in using a nonparametric hi-erarchical Bayesian model that can incorporate something like a GMM at the base level. One such model is the Hierarchical Dirichlet Process (HDP) of (Teh et al. 2006), where the hierarchy affects the weights over a collection of base models (e.g., Gaussians). Yet inference in this model only yields clustering at the base level, i.e., the channels. While such channel clustering is potentially helpful for clinicians, seizure and patient level clustering is more in line with the most important clinical questions. Ideally, we would use a model that can cluster on all three levels *at the same time*. Figure 1 depicts such a scenario for 3 toy patients. Another model, the Nested Dirichlet Process (NDP) of (Rodríguez et al. 2008), is similar to the HDP but involves multiple levels of clustering. The nesting structure of the NDP means that higher-level atoms do not share the same lower-level atoms. In our seizure data application, that would mean that seizures of a different type would share no channel information, a trait that is unrealistic for multiple seizures, which may globally be "different" but have subsets of channels that behave very similarly.

We thus present a new model that in a sense blends the desirable aspects of both the HDP and NDP models. We call this model the multi-level clustering hierarchical Dirichlet Process (MLC-HDP) because it simultaneously clusters on multiple levels of a dataset and also retains a hierarchical structure. The novel contributions of this work are:

- a model capable of clustering on multiple levels that shares lower-level atoms among higher-level atoms,

- demonstrations of how the MLC-HDP model surpasses state of the art NDP and DP models on tasks in simulated and human seizure datasets,

- the ability of the MLC-HDP model to answer previously unattempted questions in the field of quantitative epilepsy research.

## 2. Model description

Consider again the structure of our epilepsy data[2], given in the schematic in Figure 1. This data has multiple levels on which we assume clustering occurs. Each level's clustering can be thought of as a finite collection of atoms, which are multinomial at the higher levels and any arbitrary distribution at the base level. The seizure and patient levels—levels 2 and 1, respectively—are made up of multinomial atoms that represent priors over the atoms in the level below, so level-2 seizure atoms denote different seizure types, each of which is a prior over the base-atom channel-types. Level-1 patient atoms denote different patient types, each of which is a prior over the level-2 seizure-types.

---

[1] The number of "active" channels can even vary between seizures of the same patient because some channels drop in and out over time.

[2] Of course, while this model is motivated by our seizure iEEG data, it generalizes to any data where multiple levels of clustering is reasonable. Furthermore, while our model contains three levels, extending it to more or reducing it to fewer levels is straightforward.

Consider a dataset with $T$ patients, each $t$ of which has $J_t$ seizures, each $j$ of which has $N_{tj}$ channel observations, which we call $\mathbf{x}_{tji} \in \mathbb{R}^d$, where $d$ is the feature-space dimension[3]. We model the observations $\{\{\{\mathbf{x}_{tji}\}_{i=1}^{N_{tj}}\}_{j=1}^{J_t}\}_{t=1}^{T}$ of all the seizures using a set of unique base-distribution atoms with prior measure $H$ (and parameters $\boldsymbol{\lambda}$) for an arbitrary distribution $F$,

$$\mathbf{x}_{tji} \sim F(\boldsymbol{\theta}_{tji}) \qquad (1)$$

where $\boldsymbol{\theta}_{tji}$ are the parameters of the model, which are equal to those of a unique base-level atom $\boldsymbol{\phi}_k$. In the rest of this paper, $F$ is a multivariate Normal with diagonal covariance, $F(\boldsymbol{\theta}) = \mathcal{N}(\boldsymbol{\mu}, \boldsymbol{\sigma}^2)$ for $\boldsymbol{\mu} \in \mathbb{R}^d, \boldsymbol{\sigma}^2 \in \mathbb{R}_+^d$. In this paper, we will assume all priors are conjugate and so for our two-parameter Normal base model will use a Normal scaled inverse-$\chi^2$ ($\mathcal{N}$-Inv-$\chi^2$) joint prior on $\boldsymbol{\mu}$ and $\boldsymbol{\sigma}^2$ for $H$. See the Supplementary Materials for more details on the resulting posterior distribution and its sufficient statistics.

For convenience, we use an indicator variable $z_{tji}^{(3)} = k$ to describe which base-level atom models channel $i$'s activity in seizure $j$ of patient $t$. Superscripts in variables denote the data-level with which they are associated.

The set of base-atoms $\{\boldsymbol{\phi}_k\}_{k=1}^{\infty}$ have a corresponding set of stick-breaking priors (which we shall often call weights). In a 2-level HDP model of a single patient's seizures, the parent/root DP $G_0$ has its weights $\boldsymbol{\beta}$, and each seizure DP $G_j$ has its own set of weights $\boldsymbol{\pi}_j \sim \text{DP}(\alpha, \boldsymbol{\beta})$, whose posterior is a balance between the parent DP's $\boldsymbol{\beta}$ and the frequency with which the observations of seizure $j$ occur in the various base-level atoms. In our MLC-HDP version of the model, we have a set of multinomial weights atoms, $\{\boldsymbol{\pi}_\ell^{(3)}\}_{\ell=1}^{\infty}$, one of which we select to use for the base-level atom weights for seizure $j$.

Each seizure-type atom, a prior over the base atoms, is a sample from a DP,

$$\begin{aligned}\boldsymbol{\pi}_\ell^{(3)} &\sim \text{DP}(\alpha^{(3)}, \boldsymbol{\beta}^{(3)}) \\ \boldsymbol{\beta}^{(3)} &\sim \text{GEM}(\gamma^{(3)})\end{aligned} \qquad (2)$$

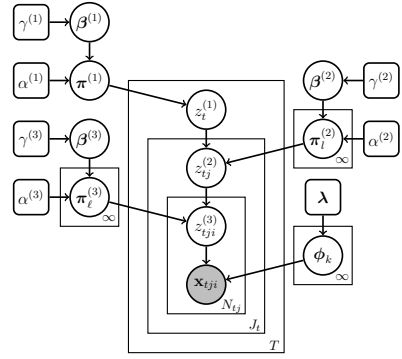

Figure 2. The directed graphical model of a 3-level MLC-HDP. The observed value is given in gray, and priors are given as squares.

where $\boldsymbol{\beta}^{(3)}$ is a distribution over the positive integers[4]. We let the indicator variable $z_{tj}^{(2)} = \ell$ denote which seizure-type patient $t$'s seizure $j$ belongs to. The patient-type atoms $\boldsymbol{\pi}_l^{(2)}$ have a similar construction as do the weights $\boldsymbol{\pi}^{(1)}$ of the patient population over the various patient types. The indicator variable $z_t^{(1)} = l$ denotes which patient-type patient $t$ belongs to.

Putting it all together, we have

$$\begin{aligned}\boldsymbol{\beta}^{(1)} &\sim \text{GEM}(\gamma^{(1)}) \\ \boldsymbol{\pi}^{(1)} &\sim \text{DP}(\alpha^{(1)}, \boldsymbol{\beta}^{(1)}) \\ z_t^{(1)} &\sim \boldsymbol{\pi}^{(1)} \\ \boldsymbol{\beta}^{(2)} &\sim \text{GEM}(\gamma^{(2)}) \\ \boldsymbol{\pi}_l^{(2)} &\sim \text{DP}(\alpha^{(2)}, \boldsymbol{\beta}^{(2)}) \\ z_{tj}^{(2)} &\sim \boldsymbol{\pi}_{l=z_t^{(1)}}^{(2)} \\ \boldsymbol{\beta}^{(3)} &\sim \text{GEM}(\gamma^{(3)}) \\ \boldsymbol{\pi}_\ell^{(3)} &\sim \text{DP}(\alpha^{(3)}, \boldsymbol{\beta}^{(3)}) \\ z_{tji}^{(3)} &\sim \boldsymbol{\pi}_{\ell=z_{tj}^{(2)}}^{(3)} \\ \boldsymbol{\phi}_k &\sim H(\boldsymbol{\lambda}) \\ \mathbf{x}_{tji} &\sim F(\boldsymbol{\phi}_{k=z_{tji}^{(3)}})\end{aligned} \qquad (3)$$

where for us $H(\boldsymbol{\lambda}) = \mathcal{N}\text{-Inv-}\chi^2(\mathbf{n}_0, \boldsymbol{\mu}_0, \boldsymbol{\nu}_0, \boldsymbol{\sigma}_0^2)$ and $F(\boldsymbol{\phi}_k) = \mathcal{N}(\boldsymbol{\mu}_k, \boldsymbol{\sigma}_k^2)$. Note that the highest level of this model is equivalent to a standard DP. We depict this model schematically by the directed graphical model in Figure 2.

---

[3]In our simulation study $d = 1$, and in our epilepsy dataset $d = 5$.

[4]GEM stands for Griffiths, Engen, and McCloskey (Pitman 2006). We say $\boldsymbol{\beta} \sim \text{GEM}(\gamma)$ if for $\beta_k' \sim \text{Beta}(1, \gamma)$, we have $\beta_k = \beta_k' \prod_{\kappa=1}^{k-1}(1 - \beta_\kappa')$ for $k = 1, \ldots, \infty$.

## 3. Model implementation

### 3.1. Notation

Our implementation is very similar to the collapsed Gibbs sampler used by (Teh et al. 2006) for the HDP[5]. For convenience, we introduce a counts variable $n$ and a sampled variable[6] $m$ at each level of the model: $n^{(3)}_{\ell k}$ denotes the number of observations in base-atom $k$ (channel-type) across the seizures in level-2 atom $\ell$, with similar meanings for $n^{(2)}_{l\ell}$ and $n^{(1)}_l$ for levels 2 (seizure types) and 1 (patient types), respectively; $m^{(3)}_{\ell k}$ denotes a sampled weight parameter on base-atom $k$ for the level-2 atom $\ell$; $m^{(2)}_{l\ell}$ and $m^{(1)}_l$ are similar. A dot ($\cdot$) in place of a subscript indicates a marginal count.

While in theory the number of atoms at each level is infinite, the number that exist at any given point is finite because the number of data points in the model is finite. We thus use $K$ to denote the current number of non-empty base-level atoms, and $L^{(2)}$ and $L^{(1)}$ to denote the current number of level-2 and level-1 atoms. For algorithmic elegance, we sometimes use $L^{(3)}$ in place of $K$. We also keep an extra, empty atom at each level to represent selecting a new atom, in which case another extra atom is appended to the list.

We explicitly sample the parameters $\phi_k = (\mu_k, \sigma^2_k)$ for our $K$ base atoms, where $\Phi_k$ describes the sufficient statistics for each base atom $k \in \{1, \ldots, K\}$ and let $\Phi_{K+1}$ describe those for new (empty) atom. The likelihood of $\mathbf{x}$ under atom $k \in \{1, \ldots, K+1\}$ is given as $f_k(\mathbf{x}) = \mathcal{N}(\mathbf{x} \,|\, \mu_k, \sigma^2_k)$.

For convenience, we denote the collection of channel-activities in a given seizure $j$ of patient $t$ as $\mathbf{s}_{tj} = \{\mathbf{x}_{tji}\}_{i=1}^{N_{tj}}$, whose joint likelihood is simply the product of the likelihoods of the individual channel-observations.

### 3.2. Markov Chain Monte Carlo Sampling

One can see from Figure 2 that there are a number of parameters of the MLC-HDP that need to be sampled. We break these variables into three main groups: the atom indicators, $z$; the level parameters, $\beta$ and $\pi$; and the base parameters, $\phi_k$. An optional fourth step is to place priors on the hyperparameters $\gamma$ and $\alpha$ and sample values for them as well. Of these steps, the first is usually the most computationally intensive. Below, we briefly summarize each of these four steps.

**Sampling atom indicators** We sample atom indicators on each of the three levels using roughly the same technique: (a) calculate the prior weights over the atoms, (b) calculate the likelihood of each data point under each of the atoms[7], (c) sample each data point's atom indicator from the posterior multinomial distribution, and (d) update the atom counts and sufficient statistics for the relevant base atoms if the atom indicator has changed. At the base level, each data point is an individual channel observation, whereas a seizure-level "data point" comprised of all the channel observations in that seizure. A patient "data point" is similarly comprised of all the seizure data points for that patient.

**Sampling level parameters** As previously described, each level can be thought of as containing a set of multinomial atoms over a set of lower-level atoms (which we call "sub-atoms"), so the seizure level contains seizure-type atoms, each of which is a particular distribution over the channel base-atoms. The patient level contains patient-type atoms, each of which is a particular distribution over the seizure-type atoms. Finally, we have a single "population" distribution over the different patient-type atoms.

For each level, we first sample the $m$ variables for each combination of sub-atom and level-atom and then use them in sampling the parent sub-atom weights, $\beta$. We could also sample the sub-atom weights, $\pi$, but again in practice use the Rao-Blackwellized approach that integrates out $\pi$.

**Sampling base parameters** Sampling the base parameters $\phi_k$ for each base atom $k$ will depend on the particular base distribution $F$ used. The Supplementary Materials give these details for our choice of a multivariate Normal with diagonal covariance. We check if any observations have been assigned to the last base atom, and if they have, we add another (empty) base atom. With no data, the last empty atom's posterior is simply the base

---

[5] Specifically, see Section 5.3

[6] Our variables $n$ and $m$ are equivalent to the number of diners eating and the number of tables serving dish $k$ in the Chinese Restaurant Process metaphor.

[7] In practice, we have found that following the method of (Teh et al. 2006) and when possible using a Rao-Blackwellized Gibbs sampler (Casella & Robert 1996; Sudderth 2006) usually leads to lower autocorrelation and slightly improved model performance. We thus generally use this method as an alternative to the likelihood under each base atom's sampled parameters. The Rao-Blackwellized sampler uses the posterior predictive likelihood instead.

Table 1. Parameters for the true distributions $p_T = \sum_i w_i \mathcal{N}(\mu_i, \sigma_i^2)$ used in the simulation study

|      | Comp 1 | | | Comp 2 | | | Comp 3 | | | Comp 4 | | |
|------|-----|---|------------|-----|------|------------|-----|-----|------------|-----|------|------------|
| Dist | $w$ | $\mu$ | $\sigma^2$ | $w$ | $\mu$ | $\sigma^2$ | $w$ | $\mu$ | $\sigma^2$ | $w$ | $\mu$ | $\sigma^2$ |
| T1   | .75 | 0 | 1.0 | .25 | 3.0 | 2.0 |     |     |     |     |      |     |
| T2   | .55 | 0 | 1.0 | .45 | 3.0 | 2.0 |     |     |     |     |      |     |
| T3   | .40 | 0 | 1.0 | .30 | -2.0 | 2.0 | .30 | 2.0 | 2.0 |     |      |     |
| T4   | .39 | 0 | 1.0 | .29 | -2.0 | 2.0 | .29 | 2.0 | 2.0 | .03 | 10.0 | 1.0 |

prior, so the parameters $\phi_k$ are sampled from the prior $H(\boldsymbol{\lambda})$.

**Sampling level hyperparameters** We sample each level's hyperparameters $\alpha$ and $\gamma$, which can have Gamma$(a,b)$ priors. This step is performed after sampling the other level parameters.

A single sampling iteration for the full MLC-HDP proceeds through sampling the atom indicators, the level parameters (and, optionally, also the hyperparameters), and the base parameters. Finally, any non-last empty base- and level-atoms are removed (and the appropriate indicator variables decremented accordingly). The Supplementary Materials give explicit algorithms corresponding to these steps[8].

## 4. Experiments

### 4.1. Simulated data

To explore some of the properties of the MLC-HDP in a controlled setting and to compare it with a similar model, we ran a 2-level version of it on the same simulated data presented in (Rodríguez et al. 2008) and implemented the Nested Dirichlet Process model described in their paper. Briefly, samples were generated from one of four distributions (T1-T4), each of which is a mixture of two to four Gaussians. The parameters of the Gaussians are given in Table 1.

We used a dataset with 5 samples from each of the four distributions, for 20 samples total. Each sample contained 100 observations from the particular distribution. We used the same hyperparameters described in (Rodríguez et al. 2008). Posterior inference for each model was run over 25 chains, each with a 5000 sample burn in and 10 sample thinning, gathering 400 samples for each chain or 10,000 samples total.

We found that the MLC-HDP gives better estimates of the four true GMM distributions than the NDP, as shown graphically on the left of Figure 3. The MLC-HDP usually found three top-level atoms, whereas the NDP balanced roughly equally between two and three (supporting the result (Rodríguez et al. 2008) give in their Figure 3 for $J = 20$ and $n = 100$). The density function estimates of both methods for T3 and T4 are thus the same, since those two true distributions are the same except for a small additional mode at $x = 10$ in the fourth distribution. The Kullback-Liebler divergence[9] of the true density function to the estimated density function for each method, shown on the right of Figure 3, also illustrates how the MLC-HDP estimates are closer to the true distributions. The major difference between the MLC-HDP and NDP models is the fact that higher-level atoms in the NDP have their own sets of base atoms, whereas higher-level atoms in the MLC-HDP share base atoms. In datasets where different group types may have observations from the same or similar base distributions (e.g., T1 and T2), the NDP must estimate those base distributions independently for each group-level atom whereas the MLC-HDP estimates benefit from data across all groups.

Another difficulty of having separate sub-atoms for each higher-level atom is that the total number of base atoms can quickly become computationally infeasible as the model is extended from two to three or more levels. For example, a modest-sized model of $K = 55$, $L^{(2)} = 35$, and $L^{(1)} = 35$ (in the NDP truncated setting) would have roughly 67,000 base atoms to sample and calculate likelihoods under, a number we have anecdotally found to be far too large for practical use.

### 4.2. Human seizures on intracranial EEG

We compiled a dataset of 193 intracranial EEG seizure records across 10 randomly-selected patients from the Children's Hospital of Philadelphia. These patients display attributes common in epilepsy datasets intracranial EEG: unique electrode placement, large discrepancies in the number of seizures per patient, and differences in the number of useable channels within

---

[8]See http://www.seas.upenn.edu/~wulsin for supplementary materials, code, and a link to the EEG dataset used.

[9]We used $D_{\mathrm{KL}}(\text{true} \| \text{estimated})$.

A Multilevel Clustering HDP Model for Human Seizures

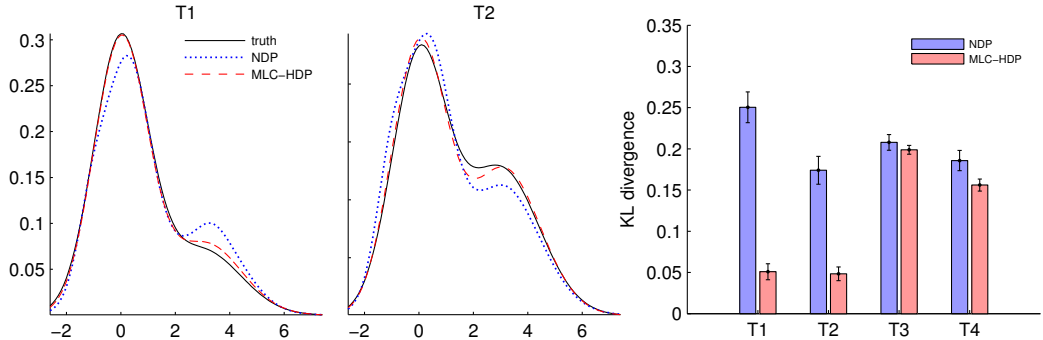

*Figure 3.* True and average estimated density functions for the T1 (**left**) and T2 (**middle**) density functions. The mean Kullback-Leibler (KL) divergence in each of the four distributions (**right**) for the NDP and MLC-HDP. Error bars denote standard error.

*Table 2.* The number of recorded seizures for each patient and whether all the seizures of each patient contained the same number of active channels.

| Patient | # Seizures | Same # channels |
|---------|------------|-----------------|
| A | 1 | yes |
| B | 9 | yes |
| C | 4 | yes |
| D | 18 | no |
| E | 61 | no |
| F | 50 | no |
| G | 1 | yes |
| H | 22 | yes |
| I | 13 | no |
| J | 14 | yes |

the seizures of a patient. Table 2 describes the number of seizures per patient as well as whether a patient's seizures contain the same number of active electrodes.

We extracted all the channel activities[10] from -30 to +90 seconds around the clinically-marked start of each seizure. We calculated a set of simple and intuitive features for each channel: the $\log_{10}$ power in four clinically relevant frequency bands (4-8, 8-13, 13-30, 30-100 Hz) for each channel over the 120 seconds, using a sliding window of 500 ms with 50% overlap. These features were chosen because they closely resemble what we believe actual epileptologists look at when reading EEG. We concatenated the four features at each of the 479 time points to get a 1916-dimensional feature vector for each channel and subsequently reduced it to 5 dimensions using PCA over all seizures of all patients[11].

---

[10] In EEG, channel activities are referential voltages, usually on the scale of mV for intracranial recordings.

[11] This reduction retained 67.7% of the original variance. Clearly, a lot of redundancy exists in the 1916-dimensional feature space for each channel.

**The advantages of a hierarchical model** In our first iEEG experiment, we examined how well the MLC-HDP model generalizes to held-out data compared with two standard Dirichlet Process (DP) models[12]. For a given patient $t$, the models were evaluated after each seizure $j$, where the seizures $j+1, \ldots, J_t$ after $j$ were used as the held out testing set. The three models and their training data are described below

**M1** The channel-observations from seizures $1, \ldots, j$ of patient $t$ are used to train a standard DP mixture model.

**M2** The channel-observations from seizures $1, \ldots, j$ of patient $t$ and all the seizures $j' \in \{1, \ldots, J_{t'}\}$ of all the other patients $t' \neq t$ are used to train a standard DP mixture model.

**M3** The same data as M2 is used but organized in the full patient-seizure-channel hierarchy available in the MLC-HDP model.

The models were evaluated using the using the conditional perplexity ($PP$) (Teh et al. 2006) of the future seizures $j+1, \ldots, J_t$ given the assigned base atoms for each channel,

$$PP(\mathbf{s}_{j+1}, \ldots, \mathbf{s}_{J_t} \mid \cdots) = \\ \exp\left(-\frac{1}{J_t-j}\sum_{j'=j+1}^{J_t}\log p\left(\mathbf{s}_{j'} \mid z^{(3)}_{tj'1}, \ldots, z^{(3)}_{tj'N_{tj'}}\right)\right) \quad (4)$$

with

$$p\left(\mathbf{s}_{j'} \mid z^{(3)}_{tj'1}, \ldots, z^{(3)}_{tj'N_{tj'}}\right) = \prod_{i=1}^{N_{tj'}} f_{(k=z^{(3)}_{tj'i})}(\mathbf{x}_{tj'i}) \quad (5)$$

Lower perplexity values indicate better models. We ran 25 chains each of the three models at each time point $j = 1, \ldots, J_t - 1$. We used used 500 samples for burn in and 20 sample spacing to get 200 samples per chain per time point in each model, or 5000 total samples per time point.

---

[12] This experiment is similar to one in (Teh et al. 2006).

The results of these experiments for patient B are shown on the left side of Figure 4. Other patients had similar plots. In the DP model with only patient B's seizures (M1), the perplexity of the first two time points is quite high since the model has only the first few seizures as training data, and the subsequent seizures are somewhat to quite different from these. The DP model (M2) with all of the other seizures as training data in addition to those of M1 performs much better in the first few time points, but as M1 has more and more training seizures, its patient-specific model becomes better than the non-specific one (M2). Though it has the exact same training and testing data as the M2 model, the MLC-HDP (M3) model consistently performs better than both M1 and M2. We believe these results show the value of employing a hierarchical model organization when it is possible and appropriate. Such an organization allows a local model (such as one for a particular patient) to get information from other, related models (such as those for other patients) without being unduly influenced by them. Characteristics of epilepsy datasets like ours, where the number of seizures can vary widely between patients, make hierarchical models like this particularly appropriate. To our knowledge, this works describes the first ever use in the epilepsy community of hierarchical models to integrate information across multiple events and subjects. This experiment shows the improvements such models can have.

**Seizure clustering performance** In our second iEEG experiment, we compared the MLC-HDP's seizure clustering (stored in the $z_{tj}^{(2)}$ indicators variables) to those of a board-certified epileptologist and those of a standard DP mixture. Comparing seizures between patients is not exactly straightforward because their channels are located on the brain in completely different configurations. Even seizures of the same patient may have different numbers of active channels from seizure to seizure as some drop in and out (see Table 2). The MLC-HDP solves this problem by defining seizure types as distributions over channel types, so different numbers of channels are accommodated simply as a different number of observations from a complex, multi-modal density (which for us is a mixture of Gaussians). For models like the DP (which is currently the only alternative for nonparametric clustering of seizures), we need features of a seizure that to not depend on the number of channels present. While various metrics have been proposed and used in the epilepsy literature, we believe the six features of (Schiff et al. 2005) capture most of the important dynamics of a seizure, namely, the synchronization of different areas of the brain and their frequency characteristics during a seizure. These features were calculated using the same 500 ms time window with 50% overlap. As with the channel features, we concatenated the six seizure features for each time point into a large vector (2874 dimensions) and then reduced them to 20 dimensions, retaining 72.3% of the variance.

Clustering seizures, even for board-certified epileptologists practicing in the same hospital, is an inherently subjective and uncertain task. We thus had two physicians cluster the same 193 seizures in our dataset (independantly from and blind to the MLC-HDP's clusterings) to get an idea of the potential variability from one human professional to another. For the sake of our subsequent analysis, we arbitrarily chose one to be the "gold standard" (though of course there is no such thing). The results we report do not substantially change when the other physician's markings were used as the standard instead. To assess the similarity between two clusterings, we used Rand's C-statistic (Rand 1971), which can accommodate different numbers and labels of clusters between two assignments of the same $N$ points.

The average similarities of the DP, MLC-HDP, and human physician seizure clusters to those of the "gold standard" clusterings are shown on the right side of Figure 4. We notice first that the two physicians usually agree most closely with each other, as expected. Second, the MLC-HDP is almost always better than the DP and in some patients is very close to the other physician. In the one patient (patient H) where the MLC-HDP performs worse than the DP, the other doctor is even farther from the "gold standard" clustering. This difference results from the methods differing in how much they split the very similar-looking seizures of patient H. The "gold standard" doctor and DP split them less, whereas the MLC-HDP and other doctor split them more.

We attribute the performance difference between the two models mostly to the fact that the DP simply has a less descriptive form of the data—one that ignores the behavior of individual channels in favor of that of the population—than the humans and MLC-HDP. Anecdotally, the physicians both remarked that they sorted the seizures by often looking at the activity of just a handful of prominent channels. This fact, along with the superior clustering results shown in Figure 4, leads us to believe that any approach to modeling seizures must begin by modeling channels and build up from there, as our MLC-HDP does. We believe that the absence of such methods until now explains the non-existence of seizure clustering in the epilepsy literature.

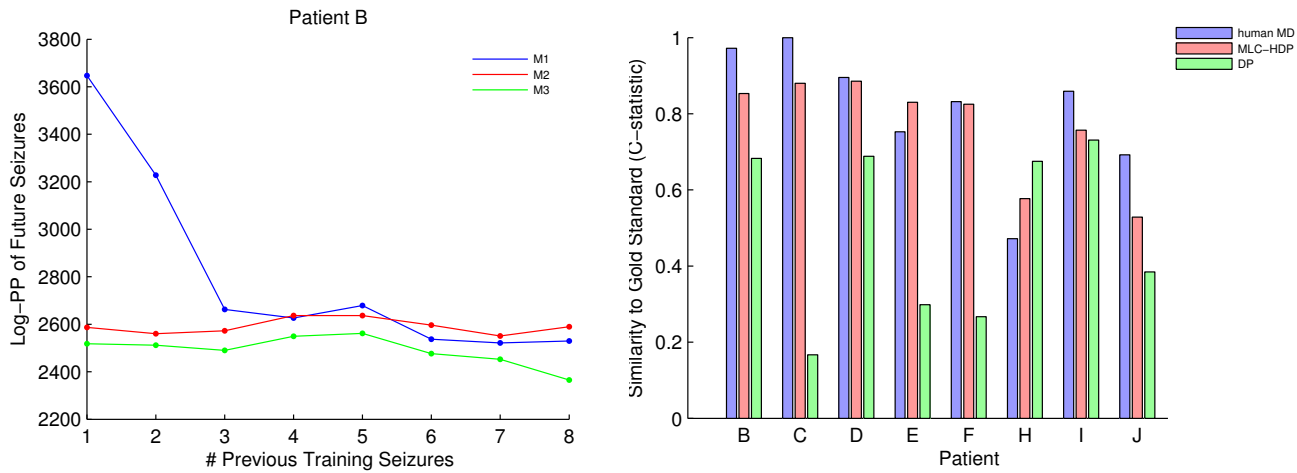

*Figure 4.* (**left:**) The mean log-perplexity of patient B's future seizures in each of the three models: M1, a standard DP with training data from previous seizures of the patient; M2, a standard DP with training data from previous seizures of the patient as well as all the seizures from all the other patients; M3, the MLC-HDP model with its full patient-seizure-channel hierarchy and clustering. (**right:**) The average seizure clustering similarity between the clusters found by a DP, an MLC-HDP, and human doctor to a second human doctor's clustering for each patient individually and for all patients together. Patient A and G are excluded because they only had one seizure. Standard error values are too small to show up on these plots.

## 5. Conclusion

We describe a new hierarchical Dirichlet Process variant, the multi-level clustering hierarchical Dirichlet Process (MLC-HDP), that simultaneously clusters on multiple levels of a dataset. In a simulation study we illustrate some advantages of this model over a similar model, the Nested Dirichlet Process. Finally, we demonstrate how the MLC-HDP allows us to build models of seizures that account for the importance of individual channels while also integrating information from many seizures within and between patients. Such a model allows us answer important clinical questions like "how many seizure types does this patient have?" and "what seizures of other patients is this seizure similar to?," questions that to the best of our knowledge have hitherto been unanswerable in the field of quantitative epilepsy analysis.

## Acknowledgements

We thank Eric Marsh and Brenda Porter, of the Children's Hospital of Philadelphia, for the continuous iEEG records, manual seizure clustering, and helpful discussion as well as Emily Fox for her helpful ideas and discussion. This work was supported by NIH grants RO1-NS041811, RO1-NS48598, and 5U24NS063930-03, the Julies Hope Award from the Citizens United for Research in Epilepsy, and the Mirowski Discovery Fund for Epilepsy Research.